\documentstyle[sprocl]{article}

\def\Journal#1#2#3#4{{#1} {\bf #2}, #3 (#4)}

\def\PRD{{\em Phys. Rev.} D}

\def\normord#1{\mathopen{\hbox{\bf:}}#1\mathclose{\hbox{\bf:}}}
\def\ha{{1\over 2}}
\def\fr#1,#2{{#1\over #2}}
\def\senk#1{\vec #1}

\def\be{\begin{equation}}
\def\ee{\end{equation}}
\def\bea{\begin{eqnarray}}
\def\eea{\end{eqnarray}}

\begin{document}

\title{REGULATING THE $P^+ = 0$ SINGULARITY}

\author{Gary McCartor}

\address{Department of Physics, SMU, 
Dallas, TX 75275, USA\\E-mail: mccartor@mail.physics.smu.edu} 




\maketitle\abstracts{ I shall discuss the regulation of the $P^+ = 0$ singularity and give some examples.  Regulating the singularity induces new operators into the theory.  This process seems rather different for the case of ultraviolet singularities than for the case of infrared singularities.}

\section{Introduction}

I want to discuss the problems involved in regulating the $p^+ = 0$ singularity, a task which always must be done when using the light-cone representation (or light-cone gauge) to study quantum field theories.  The $p^+ = 0$ point can have singularities  which are associated with both the ultraviolet and the infrared structure of the theories.  I shall give some examples of each type of singularity.  For the case of ultraviolet singularities I shall briefly discuss the case of the one loop correction to the mass in Yukawa theory.  For the case of infrared singularities I shall discuss the Schwinger model --- the best understood case.

In each case we shall see that there are issues, unfamiliar from the equal-time representation, with how to regulate the theory and with how to find the operator mixing which results from the regulation.  At the end I shall remark briefly on the import (in my view) of the examples to the problem of performing practical calculations for realistic theories in the light-cone representation.

\section{Ultraviolet Singularities}

The first example formed the starting point for the work that Stan Brodsky, John Hiller and I have been doing for some time~\cite{hi}.  I want to use it to, among other things, warn you against an argument for the equivalence of the light-cone and equal-time representations, at least at the level of perturbation theory.  The argument is sometimes given that if a finite (that is, regulated) Feynman integral is given by
\begin{equation}
            \int d^4k \quad f(k_\mu)
\end{equation}
And if
\begin{equation}
      \int dk_+ \quad  f(k_\mu) = g(k_-,k_\perp)
\end{equation}
And if light-cone perturbation theory for the same quantity leads to the integral
\begin{equation}
        \int dk_- d^2k_\perp \quad g(k_-,k_\perp)
\end{equation}
Then clearly, at least perturbatively, the two formulations are the same.

Not so~\cite{bhm}.  Let us look at an example.  For scalar Yukawa theory it is known that one Pauli-Villars field is sufficient to regulate the Fermion self-mass.  The one loop correction to the mass is easily written down as
\begin{equation}
\tilde u(p) u(p) \delta m = {-i \alpha \over 4\pi^3} \tilde u(p) \int d^4k {\gamma\cdot (p - k) +m \over[(p - k)^2 - m^2 + i\epsilon][k^2 - \mu^2]} u(p) - [\mu\rightarrow\mu_1]
\end{equation}
The integral can be evaluated in general but we need only the small-m limit, which is
\begin{equation}
     \delta m = -{3 \over 2} \alpha m \ln{\mu_1 \over \mu}
\end{equation}
We recall particularly, that the correction to the mass is zero if the bare mass is zero.  That result is true to all orders since the zero-bare-mass Fermion is protected from gaining a mass by a discrete chiral symmetry.  We can easily calculate that 
\bea
   &&(\tilde u(p) u(p))^{-1}{-i \alpha \over 4\pi^3} \tilde u(p) \int dk_+ {\gamma\cdot (p - k) +m \over[(p - k)^2 - m^2 + i\epsilon][k^2 - \mu^2]} u(p)) - [\mu\rightarrow\mu_1] = \nonumber \\
&&{{\alpha}\over {4m\pi^2}}{{p^2\vec{q}_\perp^2+(2p-q)^2m^2}\over
           {p^2\vec{q}_\perp^2+q^2m^2+p(p-q)\mu^2}} ) - [\mu\rightarrow\mu_1]
\eea
Furthermore, we find that light-cone perturbation theory gives
\be
{{\alpha}\over {4m\pi^2}} \int_{0}^{p} {{dq}\over {p(p-q)}}
     \int d^2q_\perp
     {{p^2\vec{q}_\perp^2+(2p-q)^2m^2}\over
           {p^2\vec{q}_\perp^2+q^2m^2+p(p-q)\mu^2}} - [\mu\rightarrow\mu_1]
\ee
It is easy to check that the above integral is divergent.  If we use two Pauli-Villars fields we find that the light-cone integral is now finite, but is not equal to the Feynman result and not proportional to m.  We get
\be
        \delta m = {\alpha \over 8\pi^2}\left[ \sum_{i =0}^2 \mu_i^2 \ln \mu_i^2 - m({11\over 2} + \sum_{i =0}^2 \ln \mu_i^2)\right]
\ee
If we go to three Pauli-Villars fields the light-cone integral is both finite and proportional to m. Indeed, with three Pauli-Villars fields the renormalized light-cone series is equal to the renormalized Feynman series`\cite{cy}.  

It is worthwhile understanding the source of the failure of the argument presented at the beginning of this section.  The failure of the argument is due to the meaning of the word ``convergent'' when we say that the Feynman integral is ``convergent''.  The integral is conditionally convergent and thus any value ascribed to it is a prescription.  The standard value presented above comes from the prescription: one Pauli-Villars field plus a Wick rotation.  Since the integral is conditionally convergent it is guaranteed that there exists some way of expanding the domain of integration to cover the whole space such that any value will be obtained for the integral.  The light-cone integral includes regions of the integration domain in an order not equivalent to a Wick rotation; the result might therefore be different and in this case it is.

If the ``covariant regulator'' is used, where the range of integration is limited by
\be
     {{\senk{k^2} + \mu^2}\over {q}} + {{\senk{k^2} + m^2}\over {p - q}} \le {{\Lambda^2}\over{p}}
\ee
the result for the Fermion self-mass is
\bea
&&{{\alpha}\over {8\pi^2}} \Biggl[ \left( {{\Lambda^2}\over {2}} - \mu^2 ln \Lambda^2 + \mu^2 ln \mu^2 - {{\mu^4}\over {2\Lambda^2}}\right) \nonumber \\
&& + m^2\left( 4 ln\Lambda^2 - 4 ln \mu^2 - {{11}\over{2}} + {{\mu^2}\over {\Lambda^2}} ln \Lambda^2 - {{\mu^2}\over{\Lambda^2}} ln \mu^2 + {{6\mu^2}\over{\Lambda^2}}\right)\Biggr]
\eea
The bad features of this result make it very difficult, perhaps impossible to devise an effective renormalization. These observations form the basis for the work of Brodsky, Hiller and McCartor~\cite{bhm,bhmp}.

\section{Infrared Singularities}

The example here is the Schwinger model.  The Lagrangian for the Schwinger model in light-cone gauge is
\be
 {\cal L} =  i \bar{\psi} \gamma^{\mu}
\partial_{\mu} \psi 
- \fr{1},{4} F^{\mu \nu} F_{\mu \nu} 
-  A^{\mu} {J}_{\mu} -  \lambda A^+
\ee
Notice that we have used a Lagrange multiplier, $\lambda$, to impliment the gauge choice.  The degrees of freedom in $\lambda$ are essential to the solution and the reason has to do with regulating the $p^+ = 0$ singularity.  The operator solution is~\cite{nm}
\be
     \Psi_+ = Z_+ e^{\Lambda_+^{(-)}}\sigma_+ e^{\Lambda_+^{(+)}}
\ee
\be
      \Lambda_+ = -i2\sqrt{\pi}({\eta}(x^+) + \tilde{\Sigma}(x^+,x^-))
\ee
\be
       Z_+^2 = \fr{m^2e^\gamma},{8\pi\kappa}
\ee
\be
     \Psi_- = \psi_- = Z_-e^{\Lambda_-^{(-)}}\sigma_- e^{\Lambda_-^{(+)}}
\ee
\be
        Z_-^2 = \fr{\kappa e^\gamma},{2\pi}
\ee
\be
      \Lambda_- = -i2\sqrt{\pi}\phi(x^+)
\ee
\be
      \lambda =m{\partial }_{+}({\eta }-{\phi }) 
\ee
\be
        A_+ = \fr{2},{m} \partial_+ ({\eta} + \tilde{\Sigma})
\ee
In these relations
\be
   \tilde{\Sigma}~ = ~ MASSIVE(m = \fr{e},{\sqrt{\pi}})~~PSEUDOSCALAR
\ee
\be
 \phi~ = ~CHIRAL~SCALAR~;~~ \eta ~ = ~ CHIRAL~GHOST
\ee
The chiral fields $\phi$ and $\eta$ are functions of $x^+$; they have been omitted from many previous formulations.  Together they make up $\lambda$.  All the massless fields are regulated by the Kleiber method
\be
\phi^{(+)}(x^{+})=i(4\pi )^{-\ha}\int_{0}^{\infty
}dk_{+}k_{+}^{-1}d(k_{+})\left({\rm e}^{-ik_+ x^+}-\theta 
(\kappa -k_{+})\right)
\ee
The $x^+$-dependent fields are necessary to regulate the theory.  Let us understand that point.  The singularity in the Fermi products must be
\be
\langle \Psi_+^*(x+\epsilon)\Psi_+(x)\rangle\sim \fr{1},{2\pi\epsilon^-}
\ee
But for the part independent of the chiral fields we have
\be
 \langle :e^{i2\sqrt{\pi}\tilde{\Sigma}(x + \epsilon)}::e^{-i2\sqrt{\pi}\tilde{\Sigma}(x)}:\rangle \sim e^{-2\gamma}\fr{4},{m^2} \fr{1},{\epsilon^+\epsilon^-}
\ee
For the part depending on the chiral fields we have
\be
\langle e^{i2\sqrt{\pi}\eta^{(-)}(x+\epsilon)}\sigma_+^*e^{i2\sqrt{\pi}\eta^{(+)}(x+\epsilon)}e^{-i2\sqrt{\pi}\eta^{(-)}(x)}\sigma_+e^{-i2\sqrt{\pi}\eta^{(+)}(x)}\rangle\sim e^{\gamma}\kappa \epsilon^+
\ee
Thus the chiral fields are essential for regulating the infrared singularity.  The chiral fields, necessary for regulation the theory, have other important effects.  When the completion necessary to form the full representation space is done, additional states, not in the representation space of free theory, are included.  These are translationally invariant and can mix with the vacuum; indeed gauge invariance requires that they mix with the vacuum to form a vacuum of the $\theta$-state form.  In such a state, $\bar \Psi \Psi$ has a nonzero expectation value.  

The chiral fields have no effect on the spectrum as long as the bare mass is zero.  But if the bare mass, $\mu$, is not zero, the interaction between the chiral fields and the physical fields leads to a new term in $P^-$ which does act in the physical subspace.  We find that in the physical subspace the operator is
\be
 \delta P^- \subset \mu \sigma_-^*\sigma_+ Z_-Z_+\int \normord{
e^{-i\sqrt{\pi}\tilde{\Sigma}(0,x^-)}} dx^- + C.C. \label{eq:pm}
\ee
Where the wavefunction renormalization constants have the small-$\mu$ expansions
\be
        Z_- = Z_+(\mu) =  \sqrt{\fr{\chi e^\gamma},{2\pi}} \;+\; {\cal O}(\mu)
\ee
\be
           Z_+ = Z_+(\mu) = \sqrt{\fr{m^2e^\gamma},{8\pi\chi}} \;+\; {\cal O}(\mu)
\ee
and, more generally, are determined by the relations
\be
           \{\Psi_+(x^+),\Psi_+(x^+ + \epsilon_+)\} = \delta(\epsilon_+)
\ee
\be
           \{\Psi_-(x),\Psi_-(x + \epsilon_{spacelike})\} = \delta(\epsilon_{spacelike})
\ee
This operator provides a shift in the mass-squared of the physical Schwinger particle of $2 m \mu e^\gamma cos\theta$, where $\theta$ is the vacuum angle.  In this sense the Schwinger model provides a model for the $\eta^\prime$.  It gets a mass from the anomaly which is independent of the bare mass, then has a term in the mass squared which is linear in the bare mass and linear in the chiral condensate.

Once we have identified the new term, we can drop the chiral fields and use DLCQ.  If we stay in the continuum we can calculate the renormalization constants as above (at least in principle); in DLCQ we would have to fit them to data.

\section{Remarks}

Regulating the $p^+ = 0$ singularity induces additional operators into the theory.  The form of the operators depends on the regulator used.  It is not clear to me that all regulators will allow an effective renormalization; certainly some will make renormalization much more difficult than others.  All of this is familiar from the equal-time representation.  

What is new to the light-cone representation is the unfamiliar form the new operators take and, it seems, the greater difficulty in finding them.  I am still hopeful that for the ultraviolet singularities, procedures such as using Pauli-Villars fields may provide a more or less automatic way of including the necessary operators.  I do not think that that will prove to be true in the case of infrared singularities, particularly for gauge theories.  I believe that operators similar to that of equation~(\ref{eq:pm}) will occur in four dimensions and will provide the mechanism for chiral symmetry breaking.  I do not believe that the light-quark hadrons can be correctly treated without these operators.  I do not yet know of an automatic procedure for finding them, but our knowlege of them is growing.

\section*{Acknowledgments}
This work was supported by the U.S. Department of Energy.


\section*{References}
\begin{thebibliography}{99}
\bibitem{hi}J. Hiller, {\em This Volume}

\bibitem{bhm}S. Brodsky, J. Hiller and G. McCartor, \Journal{\PRD}{58}{025005}{1998}.

\bibitem{cy}S.-J. Chang and T.-M. Yan, \Journal{\PRD}{7}{1147}{1973}.

\bibitem{bhmp}S. Brodsky, J. Hiller and G. McCartor, \Journal{\PRD}{60}{054506}{1999}.

\bibitem{nm}Y. Nakawaki and G. McCartor, {\em Prog. Theor. Phys.} {\bf 103}, 161 (2000).

\end{thebibliography}
\end{document}